\documentclass[12pt]{article}

\usepackage{latexsym,amsmath,amssymb,theorem,epsfig} 

\DeclareFontFamily{OT1}{rsfs10}{} 
\DeclareFontShape{OT1}{rsfs10}{m}{n}{ <-> rsfs10 }{} 
\DeclareMathAlphabet{\mathscript}{OT1}{rsfs10}{m}{n} 

\numberwithin{equation}{section}

\newcommand{\ns}{\normalsize}

\def\a{\alpha}
\def\b{\beta}
\def\g{\gamma}

\def\e{\epsilon}

\def\l{\lambda}

\def\s{\sigma}

\theoremstyle{plain} 

{\theorembodyfont{\rmfamily} }

\begin{document}


\begin{titlepage}

\vspace{-5cm}

\title{
  \hfill{\ns }  \\[1em]
   {\LARGE Five-Brane Dynamics and Inflation in Heterotic M-Theory} 
\\[1em] } 
\author{
   Evgeny I. Buchbinder 
     \\[0.5em]
   {\ns School of Natural Sciences, Institute for Advanced Study} \\[-0.4cm]
{\ns Einstein Drive, Princeton, NJ 08540}\\[0.3cm]}

\date{}

\maketitle

\begin{abstract}

Generic heterotic M-theory compactifications
contain five-branes wrapping non-isolated genus zero 
or higher genus curves in a Calabi-Yau threefold.
Non-perturbative superpotentials do not depend 
on moduli of such five-branes.
We show that fluxes and non-perturbative effects can stabilize them
in a non-supersymmetric AdS vacuum. 
We also show 
that these five-branes can be stabilized in a dS vacuum, if 
we modify the supergravity potential energy by Fayet-Iliopoulos terms.
This allows us to stabilize all heterotic M-theory moduli in a dS vacuum
in the most general compactification scenarios.
In addition, we demonstrate that, by this modification, one can create an inflationary 
potential. The inflationary phase is represented by 
a five-brane approaching the visible brane. 
We give a qualitative argument how extra states becoming light, 
when the five-brane comes too close, can terminate inflation. 
Eventually, the five-brane hits the visible brane and disappears 
through a small instanton transition. The post-inflationary system of moduli 
has simpler stability properties. It can be stabilized 
in a dS vacuum with a small cosmological constant.

\end{abstract}

\thispagestyle{empty}

\end{titlepage}


\section{Introduction}


Recently, there has been a considerable interest in cosmological aspects of string 
theory (see~\cite{Quevedo} for a recent review).
In~\cite{Kachru, BKQ, Becker, Raise, Vijay}, several methods
of producing de Sitter (dS) vacua in string compactifications were presented.
In~\cite{Kachru, BKQ, Raise}, it was suggested that various 
corrections to the supergravity potential energy can raise a supersymmetric
anti de Sitter (AdS) vacuum to a metastable dS vacuum. In~\cite{Becker}, based
on the earlier work~\cite{Curio},
a dS vacuum was created by balancing various exponential superpotentials
and in~\cite{Vijay}, it was argued that a dS vacuum can be created
by taking into account higher order corrections to the moduli
Kahler potential.
In addition, in~\cite{6people, Moh}, it was studied how 
effects 
of gravity and quantum particle production could trap moduli at enhanced symmetry points.
Furthermore, a substantial progress has been achieved towards inflation 
in string 
theory~\cite{Dtye, Alexander, Burg, Shiu, Choud, HHK, Raul, Dasgupta, Juan, Kallosh1, Pilo, Burg1, Oliver, Ii, Kallosh2, Kallosh3, 1, 2, 3, 4, 5}.
In these models, inflation was studied within the context of $D$-branes.
Under certain conditions the $D$-brane modulus can be treated as an inflaton.
However, all these models usually have two common problem. 
Inflation is often realized in compactification or brane world 
scenarios which do not correspond to realistic four-dimensional physics.
The other problem is that, in addition to the inflaton, there are,
usually, other moduli whose stabilization could be a problem
and whose presence can violate the slow roll conditions.

In this paper, we would like to explore the possibility of creating 
an inflationary potential within the framework 
of strongly coupled heterotic string theory, or hetorotic M-theory~\cite{HW1, HW2, Witten96}.
Such compactifications have a lot of attractive phenomenological features
(see~\cite{Faraggi} for a review on phenomenological aspects of M-theory).
Various GUT- and Standard Model-like theories were obtained 
from heterotic compactifications on Calabi-Yau threefold~\cite{DLOW, Standard, Rene, Volker}.
For example, in~\cite{Volker}, vector bundles on Calabi-Yau manifolds 
with $Z_3 \times Z_3$ homotopy group were constructed. A compactification 
on such a manifold can lead to the Standard Model with suppressed
nucleon decay. The actual particle spectrum in such theories 
was studied in~\cite{Yang1, Yang2, Yang3}. 
One more
attractive feature of such compactifications is that it is possible
to stabilize moduli in a phenomenologically acceptable range~\cite{BO, Raise}.
The set of moduli considered in~\cite{BO, Raise} was
very general. Nevertheless, it was not complete. 
In~\cite{BO, Raise}, it was assumed that the Calabi-Yau threefold had
enough isolated genus zero curves to stabilize all $h^{1,1}$ moduli. 
It was also assumed that the five-branes in the bulk wrapped only 
isolated genus zero curves. Even though such compactifications can certainly 
exist, a generic compactification with $h^{1,1}$ greater than one 
contains various, not necessarily isolated genus zero, cycles as well five-branes 
wrapped on them. In this case, it is quite possible that 
not all $h^{1,1}$ moduli can be stabilized by methods presented in~\cite{BO, Raise}.
The moduli of a five-brane wrapped on a non-isolated genus zero curve or a higher
genus curve cannot be stabilized by methods of~\cite{BO, Raise} either.
In this paper, we add these new moduli. We show that this new 
additional $h^{1,1}$ moduli can be stabilized 
in a supersymmetric AdS minimum
by the slight modification
of ideas of~\cite{BO, Raise}. 
The five-brane moduli cannot be stabilized this way. 
Surprisingly, we find that they can be stabilized 
in a non-supersymmetric AdS minimum.
Of course, what we really mean by this is that the system of moduli 
containing the moduli of this new five-brane admits a 
non-supersymmetric AdS vacuum.
However, the potential energy has one more minimum
when the five-brane coincides with the visible brane.
A heterotic M-theory vacuum can contain several five-branes 
wrapped on non-isolated genus zero or higher genus 
curves. Those which are located relatively far away from
the visible brane will be stabilized. 
On the other hand, those which are located close enough to 
the visible brane will roll towards it and, eventually, collide with it.
We show that these five-branes can be stabilized as well
by balancing the supergravity potential energy against the Fayet-Iliopoulos 
terms~\cite{DSW} induced by an anomalous $U(1)$ gauge group in the hidden sector.
This shows that the most general set of heterotic M-theory moduli can be 
stabilized. Furthermore, the cosmological constant can be positive 
and fine tuned to be very small.
By balancing the supergravity potential energy against 
Fayet-Iliopoulos terms, it is also possible 
to create a positive potential satisfying the slow roll conditions
and treat the five-brane translational modulus as an inflaton. Iflation takes place when the five-brane 
approaches the visible brane. 
However, this potential 
has one negative feature. It has a vanishing first derivative when
the five-brane coincides with the visible brane.
This means that it takes infinite time for the branes to collide.
This also means that the primordial fluctuations
will become infinite.
On the other hand, at very short distances, one cannot 
trust the low-energy field theory because new states 
are expected to become massless. At the present time,
physics at short distances in heterotic M-theory is not known.
Nevertheless, we present an argument how these new 
state can terminate inflation before the fluctuations became too big. 
Once the five-brane hits the visible brane, it gets dissolved 
into it and turns into new moduli of the vector bundle,
so called transition moduli, studied in~\cite{BDO1, BOR}.
This process is called small instanton transition~\cite{Wittensmall, Seiberg, KS, OPP}.
Thus, the post-inflationary phase does not have the inflaton but has 
extra moduli of the vector bundle. These moduli are easier to stabilize~\cite{BO}.
Therefore, the new system of moduli can be stabilized, whereas during
inflation this was not the case. In addition, 
we argue that, after a small 
instanton transition, generically, the cosmological constant changes.
It is possible to decrease the cosmological constant and, by fine tuning, 
make it consistent with observations. Let us point out that, though, besides the inflaton,
there are various other moduli during inflation, 
they are all taken into account. The potential energy has a minimum in all the other 
directions. Therefore, dynamically, one expects that all these moduli will 
roll very fast in their minimum leaving the five-brane to roll slowly 
towards the visible brane. 

This paper is organized as follows. In Section 2, we discuss the 
system of moduli in compactifications with $h^{1,1}$ greater than 
one. The reason is twofold. First, we would like 
to obtain more general results on moduli stabilization. In particular, 
we would like to stabilize the $h^{1,1}$ moduli that
do not have a non-perturbative superpotentials and, hence, cannot 
be stabilized by methods of~\cite{BO, Raise}.
Second, before we begin to study potentials for the five-brane,
whose stability properties are more complicated,
it is important to understand how the remaining moduli of the system 
are stabilized. The system of (complex) moduli considered in this section
includes the complex structure moduli, the volume modulus, two $h^{1,1}$ moduli and the moduli of the 
five-brane wrapped on an isolated genus zero cycle. 
One of the two $h^{1,1}$ moduli is assumed to be associated with an isolated 
genus zero curve and, hence, has a non-perturbative superpotential.
The other one is associated with a non-isolated genus 
zero curve or a higher genus curve. The non-perturbative 
superpotential does not depend on this modulus.
By the slight modification of ideas of~\cite{BO, Raise}, we show that
this system can be stabilized in a supersymmetric AdS vacuum.
The crucial moment is that, if $h^{1,1}$ is greater that one, it is possible to 
choose one of the contributions to the tension of the hidden brane 
to be positive without having the gauge coupling constant 
stronger in the visible sector.
The system of moduli can be supplemented by vector bundle moduli~\cite{BOR}.
They can be stabilized as well~\cite{BO}.
For simplicity, we will ignore them.
In Section 3, we add one more five-brane to the system.
This five-brane is wrapped on a non-isolated genus zero or higher genus 
curve. 
Approximately, we can treat the rest of the moduli
fixed and consider an effective potential for the remaining five-brane modulus.
This potential is very difficult to analyze analytically.
A graphical analysis shows that, generically, 
it has a non-supersymmteric AdS minimum. 
Nevertheless, if a five-brane was originally located close
to the visible brane, it will roll towards it.
In the rest of the paper, we concentrate only on 
dynamics of rolling five-branes.
We modify this effective potential with Fayet-Iliopoulos terms and show that addition of a 
Fayet-Iliopoulos term in the hidden sector can stabilize a rolling five-brane.
We also show that the cosmological constant in such a vacuum can be 
positive and small. The results from Sections 2 and 3 provide 
stabilization of heterotic moduli in the most general set-up
in a vacuum with a positive cosmological constant.
They also indicate that it is conceivable to obtain
two distinct dS vacua. One of them is the lift of the non-supersymmetric AdS minimum.
The other one is created by addition of a Fayet-Iliopoulos
term in the hidden sector. 
In Section 4, we argue that, by balancing the supergravity 
potential energy and Fayet-Iliopoulos terms, it is 
also possible to construct a potential with inflationary 
properties. One of the slow roll parameters turns out to be naturally 
much less than one. The other one can be (not necessarily fine) 
tuned to be much less than one. 
We also discuss the amount of inflation and primordial fluctuations. 
In the last subsection of Section 4, we discuss how the system 
can escape from inflation at very short distances. 
We give a qualitative argument how the appearance of new light
states in the field theory can provide such an escape.
In Section 5, we discuss the post-inflationary phase. 
After inflation, the five-brane hits the visible brane and 
disappears through a small instanton transition. 
The new system of moduli does not contain the five-brane but has extra vector
bundle moduli. Unlike the five-brane modulus, these moduli can be 
stabilized. We show that the new system of moduli can be 
stabilized in a vacuum with a positive cosmological constant which can be 
fine tuned to be very small.



\section{Supersymmetric AdS Vacua in Models with $h^{1,1}>1$}


\subsection{The System of Moduli}


In this paper, we work in the context of strongly coupled heterotic string 
theory~\cite{HW1, HW2} compactified on a Calabi-Yau threefold~\cite{Witten96, LOW4}.
To one of the orbifold fixed planes we will refer as to the visible brane 
(or the visible sector). To the other one we will refer as to the hidden brane
(or the hidden sector). Such compactifications also allow five-branes 
wrapped on holomorphic cycles in the Calabi-Yau manifold and parallel 
to the orbifold fixed planes.
Moduli stabilization in this theory 
was performed in a relatively general setting in~\cite{BO, Raise}.
Nevertheless, it was assumed in~\cite{BO, Raise} that the Calabi-Yau manifold has enough 
isolated genus zero curves to stabilize all the $h^{1,1}$ and five-brane moduli.
In this paper, we would like to consider a more complicated 
set-up when the Calabi-Yau threefold has two-cycles, one represented 
by isolated genus zero curve and the other one by curves of a different type.
They could be either non-isolated genus zero curves or curves 
of a higher genus. In both of these two cases, no non-perturbative 
superpotential for the corresponding $h^{1,1}$ modulus can be 
generated by string or, more precisely, open membrane
instantons\footnote{The statement that strings on non-isolated genus zero 
curves do not contribute to the non-perturbative superpotential was conjectured 
by Witten~\cite{Private}. The author is very grateful to Edward Witten 
for discussions on this issue.}~\cite{Private, DSWW2}. 
This also means that one cannot generate a non-perturbative 
superpotential for moduli of a five-brane wrapped on such a cycle. 
However, compactifications on a Calabi-Yau manifold with $h^{1,1}$ represented only by isolated 
genus zero cycles are, of course, very restrictive.
A generic compactification scenario involves 
a Calabi-Yau manifold with non-isolated genus zero or higher genus cycles 
and five-branes wrapped on such cycles. As an example, consider a Calabi-Yau manifold 
elliptically fibered over the Hirzebruch surface ${\mathbb F}_r, r=0, 1, \dots$. 
The Hirzebruch surface ${\mathbb F}_r$ is a ${\mathbb P}^1$ bundle 
over ${\mathbb P}^1$. We denote the class of the base of this bundle by ${\cal S}$ and the 
class of the fiber by ${\cal E}$. These Calabi-Yau threefolds are simply connected and
admit a (generically unique) global holomorphic section which we denote by $\s$.
For such a manifold, generically, we have 
\begin{equation}
h^{1,1}=3
\label{1.1}
\end{equation}
and the basis of curves can be chosen to be 
\begin{equation}
\s \cdot \pi^*{\cal S}, \quad \s \cdot \pi^*{\cal E}, \quad F.
\label{1.2}
\end{equation}
Here $\pi$ is the projection map from the threefold onto the base ${\mathbb F}_r$
and $F$ is the class of the elliptic fiber. 
The curves $\s \cdot \pi^*{\cal S}$ and $\s \cdot \pi^* {\cal E}$ have genus zero. The curve $F$
has genus one. The curve 
$\s \cdot \pi^*{\cal S}$
has a self-intersection $-r$ and, thus, is an isolated genus zero curve for $r>0$.
It is a non-isolated genus zero curve for $r=0$. The curve $\s \cdot \pi^*{\cal S}$
has a self-intersection zero for any $r$ and, thus, is non-isolated genus zero curve.
Therefore, it is important to stabilize the $h^{1,1}$ moduli corresponding 
to non-isolated genus zero or higher genus curves.
It is also important to understand whether or not it is possible 
to stabilize five-branes wrapped on such cycles.
In this paper, for simplicity, we consider the case $h^{1,1}=2$. 
We will assume that there is one isolated genus zero curve and one 
curve of a different type. The generalization to the case involving 
many curves isolated curves of various types is conceptually straightforward
but technically more difficult.

At this point, we would like to make a remark. 
It may happen that
the pullback of more than one 
harmonic form $\omega_{I}$ onto a given isolated curve is non-zero.
As a result, the non-perturbative superpotential
associated with this isolated curve may depend on the 
linear combination of more than one $h^{1,1}$ modulus. In particular, it may depend 
on all $h^{1,1}$ moduli.
In this case, all $h^{1,1}$ moduli can be stabilized by 
methods presented in~\cite{BO}. However, one might expect
that, generically, there can be $h^{1,1}$ moduli of two sorts,
those that appear in the non-perturbative superpotential and those 
which do not. In this paper, we simply assume that our compactification has
one modulus of each sort.
In this section, we will not consider five-branes wrapped 
on non-isolated genus zero or higher genus curves. We will add 
such a five-brane in the next section.

The system of moduli that we would like to consider in this section includes 
the following complex moduli
\begin{equation}
S, T^1, T^2, {\bf Y}, Z_{\a}.
\label{1.3}
\end{equation}
The modulus $S$ is related to the volume of the Calabi-Yau manifold
\begin{equation}
S=V+i\s,
\label{1.4}
\end{equation}
where $\s$ is the axion. The moduli $T^1$ and $T^2$ are 
the $h^{1,1}$ moduli. They are defined as follows~\cite{LOW4, LOSW5}
\begin{equation}
T^I=R b^I+ip^I, \quad I=1, 2.
\label{1.5}
\end{equation}
where $R$ is the size of the eleventh dimension,
$b^I$ are the Kahler moduli of the Calabi-Yau threefold
and $p^I$'s come from the components of the M-theory three-form 
$C$ along the interval and the Calabi-Yau manifold. 
The moduli $b^I$ are not all independent. They satisfy the constraint
\begin{equation}
\sum_{I,J,K=1}^{2}d_{IJK}b^I b^J b^K =6,
\label{1.6}
\end{equation}
where coefficients $d_{IJK}$ are the Calabi-Yau intersection numbers
\begin{equation}
d_{IJK}=\frac{1}{V}\int_{CY} \omega_I \wedge \omega_J \wedge \omega_K.
\label{1.7}
\end{equation}
The constraint~\eqref{1.6} reduces the number of independent $b$-moduli by one.
We will take $T^1$ to correspond to the area of the isolated genus zero curve and 
$T^2$ to the area of the remaining curve.
${\bf Y}$ is the modulus of the five-brane wrapped on the isolated genus zero curve. 
In this case, there is only one five-brane modulus~\cite{Five}, whose
real is the position of the five-brane in the bulk
\begin{equation}
{\bf Y}=y+i(a+\frac{p_1}{Rb_1}),
\label{1.8}
\end{equation}
where $a$ is the axion arising from dualizing the three-form field strength
propagating on the five-brane world-volume.
At last, by $Z_{\a}$ we denote the complex structure moduli. The 
actual number of them is not relevant for us. 
A generic heterotic compactification contains also instanton moduli~\cite{BOR}.
Their stabilization was considered in~\cite{BO}. In this section, for simplicity, 
we will ignore them. They can be added and treated as in~\cite{BO}.
However, we will come back to them in the last section.
The moduli $V, T^1, T^2$ and $y$ are assumed to be dimensionless
normalized with respect to the following reference scales
\begin{equation}
v_{CY}^{-1/6} \approx 10^{16} GeV, \quad (\pi \rho)^{-1} \approx 10^{14}-10^{15} GeV.
\label{1.9}
\end{equation}
In order to obtain the four-dimensional coupling constants in the correct phenomenological range
\cite{Witten96, Banks}, 
the corresponding moduli should be stabilized at (or be slowly rolling near) the values 
\begin{equation}
V \sim 1 \quad R \sim 1.
\label{1.10}
\end{equation}

The Kahler potential for this system is as follows \cite{Candelas, LOW4, DerS}
\begin{equation}
\frac{K}{M^{2}_{Pl}} = K_Z + K_{S,T^1, T^2, {\bf Y}},
\label{1.11}
\end{equation}
where 
\begin{equation}
K_Z= -\ln(-i \int \Omega \wedge \bar \Omega), 
\label{1.12}
\end{equation}
and 
\begin{equation}
K_{S,T,{\bf Y}} =-\ln(S+\bar S) -
\ln (d_{IJK}(T^I+\bar T^I)(T^J+\bar T^J)(T^K+\bar T^K)) +
2 \tau_5 \frac{({\bf Y}+\bar{\bf Y})^2}{(S+\bar S)(T^1+\bar T^1)}.
\label{1.13}
\end{equation}
Here $M_{Pl}$ is the four-dimensional Planck scale and $\tau_5$ is given by 
\begin{equation}
\tau_5 =\frac{T_5 v_5 (\pi \rho)^2}{M^{2}_{Pl}},
\label{1.14}
\end{equation}
where $v_5$ is the area of the cycle on which the five-brane is wrapped and $T_5$ is 
\begin{equation}
T_5 =(2 \pi)^{1/3} (\frac{1}{2 \kappa_{11}^2})^{2/3}, 
\label{1.15}
\end{equation}
with $\kappa_{11}$ being the eleven-dimensional gravitational coupling constant. It is
related to the four-dimensional Planck 
mass as 
\begin{equation}
\kappa_{11}^2=\frac{\pi \rho v_{CY}}{M^2_{Pl}}.
\label{1.16}
\end{equation}
Evaluating $\tau_5$ by using \eqref{1.16} and \eqref{1.9} gives 
\begin{equation}
\tau_5 \approx \frac{v_5}{v_{CY}^{1/3}}.
\label{1.17}
\end{equation}
Generically this coefficient is of order one. 

The superpotential for this system consists of three different contributions 
\begin{equation}
W=W_{f}-W_{g} -W_{np}.
\label{1.18}
\end{equation}
$W_f$ is the flux-induced superpotential \cite{GVW, Berndt, Constantin}
\begin{equation}
W_f =
\frac{M^2_{Pl}}{v_{CY}\pi \rho} \int dx^{11} \int_{CY} G \wedge \Omega,
\label{1.19}
\end{equation}
where $G$ is the M-theory four-form flux. 
The order of magnitude of $W_f$ was estimated in \cite{BO} and was found to be, generically, of order $10^{-8}M_{Pl}^3$.
In fact, this is flexible. The superpotential $W_f$ may receive certain higher order corrections from Chern-Simons 
invariants. In~\cite{GKLM} it was argued that these Chern-Simons invariants can reduce the order of magnitude of
$W_f$. 

By $W_g$ we denote the superpotential induced by a gaugino condensate in the hidden sector 
\cite{DRSW, Horava, LOW, Nonstandard}.
A non-vanishing
gaugino condensate has important phenomenological consequences.
Among other things, it is responsible for 
supersymmetry breaking in the hidden sector. When that symmetry 
breaking is transported to the observable brane, it leads to 
soft supersymmetry breaking terms
for the
gravitino, gaugino and matter fields \cite{KL, BIM, NOY, LT}.
See~\cite{Nil} for a good review on gaugino condensation in string theory.
This superpotential has the following structure 
\begin{equation}
W_{g} = h M^{3}_{Pl} exp(-\e S +\e \alpha_1^{(2)} T^1 +
\e \alpha_2^{(2)} T^2 
- \e \b \frac{{\bf Y}^2}{T^1}).
\label{1.20}
\end{equation}
The order of magnitude of $h$ is approximately $10^{-6}$ \cite{LOW}. The coefficient $\e$ is related to the 
coefficient $b_0$ of the one-loop beta-function and is given by 
\begin{equation}
\e = \frac{6 \pi}{b_0 \alpha_{GUT}}.
\label{1.21}
\end{equation}
For example, for the $E_8$ gauge group $\e \approx 5$.
The coefficients $\alpha_I^{(2)}$ represents the 
tension (up to the minus sign) of the hidden brane
measured with respect to the Kahler form $\omega_I$ 
\begin{equation}
\alpha_I^{(2)} =
\frac{\pi \rho}{16 \pi v_{CY}} 
(\frac{\kappa_{11}}{4 \pi})^{2/3} 
 \int_{CY} \omega_I \wedge (tr F^{(2)} \wedge F^{(2)} - \frac{1}{2} tr {\cal R} 
\wedge {\cal R}),
\label{1.22}
\end{equation}
where $F^{(2)}$ is the curvature of the gauge bundle on the hidden brane. 
Similarly, the coefficient $\beta$ is the tension of the five-brane. It is given by \cite{Visible}
\begin{equation}
\beta=\frac{2 \pi^2 \rho}{v_{CY}^{2/3}} (\frac{\kappa_{11}}{4 \pi})^{2/3}
\int_{CY} \omega_{1} \wedge {\cal W},
\label{1.23}
\end{equation}
where ${\cal W}$ is the four-form Poincare dual to the holomorphic curve on which
the five-brane is wrapped. 
Generically both $\alpha_I^{(2)}$ and $\beta$ are of order one. 
In fact, from eqs.~\eqref{1.14}, \eqref{1.15} and and \eqref{1.23} it follows that
\begin{equation}
\b \approx \tau_5.
\label{1.24}
\end{equation}
Let us note the following fact which will be crucial for stabilization of $T^2$.
If $h^{1,1}=1$, apparently, it is important to have $\alpha^{(2)}$ positive
(and, correspondingly, the tension negative) in the hidden sector. 
This, in particular, happens when the bundle on the hidden brane is trivial.
The reason is that the quantity
\begin{equation}
Re(S-\sum_I \a_I^{(2)}T^I +\b \frac{{\bf Y}^2}{T^1})
\label{1.25}
\end{equation}
represents the inverse square of the gauge coupling constant 
in the hidden sector $\frac{1}{g^2_{hidden}}$. Furthermore, 
the quantity
\begin{equation}
Re(S-\sum_I \a_I^{(1)}T^I +T^1(1-\b \frac{{\bf Y}^2}{T^1}))
\label{1.26}
\end{equation}
represents the inverse square of the gauge coupling constant 
in the visible sector $\frac{1}{g^2_{visible}}$.
Here $\alpha^{(1)}$ is the tension (up to the minus sign) of the visible brane
\begin{equation}
\alpha^{(1)} = 
\frac{\pi \rho}{16 \pi v_{CY}} 
(\frac{\kappa_{11}}{4 \pi})^{2/3} 
\int_{CY} \omega \wedge (tr F^{(1)} \wedge F^{(1)} - \frac{1}{2} tr {\cal R} 
\wedge {\cal R}),
\label{1.27}
\end{equation}
where $F^{(1)}$ is the curvature of the gauge bundle on the visible brane.
If, for example, $h^{1,1}=1$ and there are no five-branes,
we have 
\begin{equation}
\frac{1}{g^2_{hidden}}=Re(S-\a^{(2)}T)
\label{1.28}
\end{equation}
and
\begin{equation}
\frac{1}{g^2_{visible}}= Re(S-\a^{(1)}T)
\label{1.29}
\end{equation}
The anomaly cancellation condition in the absence of five-branes,
\begin{equation}
c_2(V_{visible})+c_2(V_{hidden})=c_2(TX),
\label{1.30}
\end{equation}
sets 
\begin{equation}
\a^{(2)}=-\a^{(1)}.
\label{1.31}
\end{equation}
Now it is clear that if $\a^{(2)} <0$, the gauge coupling 
constant in the hidden sector is weaker that the gauge coupling 
constant in the visible sector and the whole assumption about the gaugino 
condensation in the hidden sector breaks down.
It is unlikely that this statement changes when five-branes are included.
However, when $h^{(1,1)}$ is greater than zero, there is nothing 
wrong with having some $\a_I^{(2)}$'s negative. 
It is still possible to keep the gauge coupling constant
stronger in the hidden sector. We will assume that 
\begin{equation}
\a_1^{(2)} >0
\label{1.32}
\end{equation}
and 
\begin{equation}
\a_2^{(2)} <0.
\label{1.33}
\end{equation}
It is important to note that
the quantity given by eq.~\eqref{1.25} must be positive. 
This means that the superpotential~\eqref{1.20} cannot be 
trusted for large values of the interval size $R$. 
One should expect that higher order corrections to the combination~\eqref{1.25}
will make the gauge coupling constant $\frac{1}{g^2_{hidden}}$
well defined for large values of $R$. Partial support for this comes
from~\cite{Curio1, Curio2}. 

The last contribution to the superpotential that we have to discuss is the non-perturbative superpotential
$W_{np}$ 
\cite{DSWW1, DSWW2, Wittensuper, BBS, Witten00, Lima1, Lima2, Moore, BDO2, BDO3}.
Such a superpotential is induced by open membranes wrapped on an isolated genus 
zero curve. Therefore, it depends on the $h^{1,1}$ modulus $T^1$ and on the 
five-brane modulus ${\bf Y}$. However, it does not depend on the $h^{1,1}$ modulus $T^2$.
The non-perturbative superpotential has three parts
\begin{equation}
W_{np}=W_{vh}+W_{v5} +W_{5h}.
\label{1.34}
\end{equation}
$W_{vh}$ is induced by a membrane stretched between the visible and the hidden branes. 
It behaves as
\begin{equation}
W_{vh} \sim e^{-\tau T^1}
\label{1.35}
\end{equation}
$W_{v5}$ is induced by a membrane stretched between the visible brane and the five-brane. It behaves as
\begin{equation}
W_{v5} \sim e^{-\tau {\bf Y}}.
\label{1.36}
\end{equation}
At last, $W_{5h}$ is induced by a membrane stretched between the five-brane and the hidden brane. It behaves as
\begin{equation}
W_{5h} \sim e^{-\tau (T^1-{\bf Y})}.
\label{1.37}
\end{equation}
The coefficient $\tau$ is given by~\cite{Lima1, Lima2} 
\begin{equation}
\tau =\frac{1}{2} (\pi \rho) v_i (\frac{\pi}{2 \kappa_{11}})^{1/3},
\label{1.38}
\end{equation}
where $v_i$ is the reference area of the isolated curve. 
Generically, $\tau$ is much bigger than one.
%
%
As in~\cite{BO, Raise},
we will assume that the five-brane is close to the hidden sector.
It was argued in~\cite{BO, Raise} that only in this case it is possible 
to stabilize the size of the interval in a phenomenologically acceptable range.
Therefore, the contributions $W_{vh}$ and $W_{v5}$ decay very fast and
we have 
\begin{equation}
W_{np} = W_{5h} = M^{3}_{Pl} a e^{-\tau (T^1-{\bf Y})}.
\label{1.39}
\end{equation}
For concreteness we assume that the coefficient $a \sim 1$. 
%
%
%
%


\subsection{Supersymmetric AdS vacua}


In this subsection, we will argue that this system of moduli has an AdS
minimum. The consideration is, somewhat, similar to~\cite{BO, Raise} and we will be 
relatively brief. Let us first discuss the imaginary parts of the moduli.
A consideration analogous to~\cite{BO} shows that the imaginary parts of $T^1$ and ${\bf Y}$ are stabilized at 
values
\begin{equation}
ImT^1 \sim \frac{1}{\tau} \approx 0, \quad Im{\bf Y} \approx 0.
\label{1.40}
\end{equation}
The imaginary part of the linear combination 
$S-\alpha_2^{(2)}T^2$ is stabilized in such a way that the superpotentials 
$W_f$ and $W_g$ are out of phase. Similarly, $W_f$ and $W_{np}$ are also out phase.
We already took this into account in eq.~\eqref{1.18}
by putting the minus sign in appropriate places.
Unfortunately, the superpotential of the form~\eqref{1.18}, \eqref{1.19}, \eqref{1.20} 
and~\eqref{1.39} does not allow us to stabilize the remaining linear combination 
of $S$ and $T^2$. It can be shown to be a flat direction. This problem 
cannot be resolved even by considering a multiple gaugino condensation 
in the hidden sector. Nevertheless, it is easy to realize that 
this problematic linear combination
can be stabilized by taking into account the higher order $T$-corrections
to the gauge coupling in the hidden sector. We will make a more detailed comment
on it later in this subsection. Therefore, stabilization of the remaining
imaginary part does not represent a conceptual problem.
In the rest of the paper, we will concentrate only on the real parts 
of the moduli ignoring their imaginary parts.
Now let us consider the real parts of the moduli and show that 
the system under study indeed has an AdS minimum satisfying
\begin{equation}
D_{all{\ }fields}W=0,
\label{1.41}
\end{equation}
where $D$ is the Kahler covariant derivative.
We will not distinguish between the superpotentials and their absolute values.
First, we consider equations
\begin{equation}
D_{Z_{\a}}W=0.
\label{1.42}
\end{equation}
Assuming that 
\begin{equation}
W_f >>W_g, W_{np}
\label{1.43}
\end{equation}
in the interesting regime, eq.~\eqref{1.43}
can be written as
\begin{equation}
\partial_{Z_{\alpha}}W_{f} + \frac{\partial K_{Z_{\alpha}}}{\partial Z_{\alpha}} W_{f} =0.
\label{1.44}
\end{equation}
In \cite{BO}, it was shown that ineq.~\eqref{1.43} is indeed satisfied. In eq.~\eqref{1.44}, 
all quantities depend on the complex structure moduli only. We will assume that this equation 
fixes all the complex structure moduli. Partial evidence that equations of the type \eqref{1.44}
fix all the complex structure moduli comes, for example, from~\cite{Schulz1}. 
The next equation to consider is 
\begin{equation}
D_{S}W=0.
\label{1.45}
\end{equation}
By using eqs.~\eqref{1.13}, \eqref{1.20} and~\eqref{1.43}, it can be written as 
\begin{equation}
\e W_g =F_1W_f,
\label{1.46}
\end{equation}
where 
\begin{equation}
F_1=\frac{1}{2V}(1+\tau_5 \frac{y^2}{(Rb^1)^2}).
\label{1.47}
\end{equation}
By using eqs.~\eqref{1.13}, \eqref{1.20}, \eqref{1.39}, \eqref{1.43} and~\eqref{1.46}, 
eq.
\begin{equation}
D_{T^1}W=0
\label{1.48}
\end{equation}
can be rewritten as
\begin{equation}
\tau W_{np}=((\a_1^{(2)}+\b\frac{y^2}{(Rb^1)^2})F_1+F_2)W_f,
\label{1.49}
\end{equation}
where
\begin{equation}
F_2=\frac{3\sum_{IJ}d_{1IJ}b^Ib^J}{R\sum_{IJK}d_{IJK}b^Ib^Jb^K}
+\frac{\tau_5 y^2}{V(Rb^1)^2}.
\label{1.50}
\end{equation}
Now let us consider eq.
\begin{equation}
D_{T^2}W=0.
\label{1.51}
\end{equation}
Note that the non-perturbative superpotential~\eqref{1.39} does not depend on $T^2$,
thus, $T^2$ cannot be stabilized by the same mechanism as $T^1$. By using
eqs.~\eqref{1.13}, \eqref{1.20}, \eqref{1.43}, we obtain
\begin{equation}
\e W_g =F_3 W_f,
\label{1.52}
\end{equation}
where
\begin{equation}
F_3=-\frac{3\sum_{IJ}d_{2IJ}b^Ib^J}{\a_2^{(2)}R\sum_{IJK}d_{IJK}b^Ib^Jb^K}.
\label{1.53}
\end{equation}
Eqs.~\eqref{1.46} and~\eqref{1.52} are consistent only if 
$F_1$ and $F_3$ are equal to each other. In particular, they must have 
the same sign. This is possible only if $\a_2^{(2)}$ is negative.
As we argued before, this does not lead to any contradictions.
Note, that $F_1$ and $F_3$ are both real. This is the reason why 
only one linear combination of the imaginary parts of $S$ and $T^2$ moduli can be 
stabilized. On the other hand, if higher order $T$-corrections to the 
quantity~\eqref{1.25} are present, $F_3$ is really complex and, hence,
the imaginary parts of both $S$ and $T^2$ moduli can be stabilized.
The last equation to consider is 
\begin{equation}
D_{{\bf Y}}W=0.
\label{1.54}
\end{equation}
By using eqs.~\eqref{1.13}, \eqref{1.20}, \eqref{1.39}, \eqref{1.47} and~\eqref{1.50},
we obtain
\begin{equation}
-(\a_1^{(2)}+\b\frac{y^2}{(Rb^1)^2})F_1+F_3+
2\b\frac{y}{Rb^1}F_1 +2 \tau_5\frac{y}{VRb^1}=0.
\label{1.55}
\end{equation}
Eqs.~\eqref{1.46}, \eqref{1.49}, \eqref{1.52} and \eqref{1.55} are 
the four equations with four independent variables $V, R, y$ and one of two $b^I$'s.
Equations of this type were analyzed in detail in~\cite{BO, Raise} 
in the case of only one $h^{1,1}$ modulus. 
It was shown that they admit a solution
with the following properties
\begin{itemize}

\item 
V is of order one.

\item 
R is of order one.

\item
The gauge coupling constant $g^2_{hidden}$ does not become imaginary.

\item
The five-brane is close to the hidden brane ($R-y \sim 0.1$).

\end{itemize}
In this paper, we will not perform a detailed analysis. 
Let us just point out that in eqs.~\eqref{1.46} and~\eqref{1.49},
the run-away moduli are stabilized by fluxes.
Eqs.~\eqref{1.46} and~\eqref{1.52} lead to eq.
\begin{equation}
F_1=F_3,
\label{1.56}
\end{equation}
which is well defined if $\a_2^{(2)}$ is negative. 
Eq.~\eqref{1.55} is a purely algebraic equation.
It is possible to show that it admits a numeric solution with the right properties
as in~\cite{BO, Raise}. We will not give a numeric 
result in this paper. See~\cite{BO, Raise} for a detailed 
analysis of similar equations. 

In this subsection, we have provided stabilization of moduli listed in~\eqref{1.3}.
This list includes the modulus $T^2$, corresponding to the area of a non-isolated 
genus zero curve or a curve of a higher genus. Stabilization
of such a modulus differs from stabilization of the modulus $T^1$, corresponding
to the area of an isolated genus zero curve. The crucial point 
in stabilization of $T^2$ is that, in the case when $h^{1,1}>1$, it possible 
to choose the coefficient $\a_2^{(2)}$ to be negative. 
It is not possible to do in the case when $h^{1,1}=1$, because 
it would follow that the gauge coupling coupling in the hidden sector
became weaker than in the visible sector. This would not be consistent
with the assumption about gaugino condensation in the hidden sector.

The AdS vacuum constructed in this section can be raised to a metastable 
dS vacuum along the lines of~\cite{Raise}. 
This can be achieved by either adding Fayet-Iliopoulos terms to the supergravity 
potential energy or by working within the context of $E_8 \times \bar E_8$
theory. 


\section{Addition of a Five-Brane and dS Vacua}


\subsection{Effective Potential for a Five-Brane Modulus and Non-Su\-per\-sym\-met\-ric AdS Vacua}


Now we would like to see what happens if we add a five-brane wrapped on a non-isolated 
genus zero curve or on a higher genus curve to the system of moduli considered above.
We will denote the complex five-brane modulus by ${\bf X}$
and its real part by $x$. The Kahler potential~\eqref{1.13} receives the contribution
\begin{equation}
\Delta K=2 \tau_5^{'} \frac{({\bf X}+\bar{\bf X})^2}{(S+\bar S)(T^2+\bar T^2)}.
\label{2.1}
\end{equation}
The gaugino condensate superpotential gets modified and becomes
\begin{equation}
W_g \to W_ge^{-\e \b^{'} \frac{{\bf X}^2}{T^2}}.
\label{2.2}
\end{equation}
The coefficients $\tau_5^{'}$ and $\b^{'}$ are given by expressions similar 
to eqs.~\eqref{1.14} and~\eqref{1.23}.
Unfortunately, if a five-brane wraps a non-isolated cycle, one should expect 
other five-brane moduli in addition to ${\bf X}$. Such moduli have 
never been considered in the literature in detail. Nevertheless,
one should expect that
the gaugino condensate superpotential~\eqref{2.2} depends on them.
This might provide their stabilization.
Thus, we will assume that these 
additional moduli are fixed and not consider them in this paper. 
In principle, one can avoid this issue by taking a five-brane
wrapping an isolated higher genus curve.
Let us first see if we can stabilize ${\bf X}$ in an AdS vacuum. 
By using eqs.~\eqref{2.1} and~\eqref{2.2},
the
equation
\begin{equation}
D_{{\bf X}}W=0
\label{2.3}
\end{equation}
can be written as
\begin{equation}
\e\b^{'} \frac{{\bf X}}{T^2}W_g +\tau_5^{'} \frac{x}{VRb^2}W_f=0.
\label{2.4}
\end{equation}
It is easy to realize that the only solution for ${\bf X}$ is 
\begin{equation}
{\bf X}=0.
\label{2.5}
\end{equation}
%
The point $x=0$ corresponds to the five-brane coinciding with the visible brane.
Such a vacuum is unstable in the sense that the five-brane 
will disappear through a small instanton transition~\cite{Seiberg, KS, OPP, BDO1}
and turn into new vector bundle moduli. 

\begin{figure}
\epsfxsize=4in \epsffile{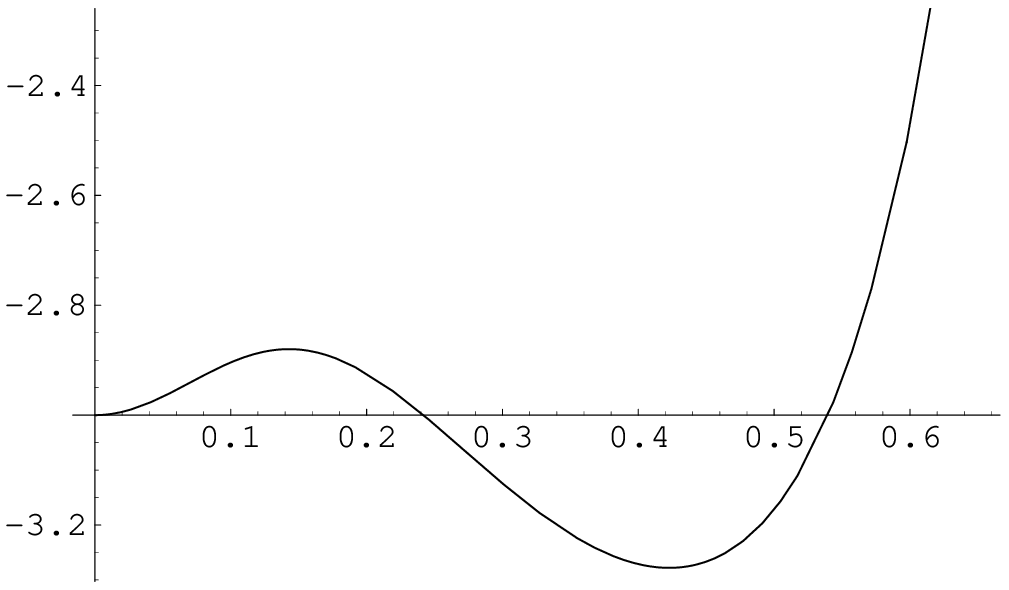}
\begin{picture}(30,30)
\put(-260,180){$\frac{U(x)}{U_0}$}
\put(5, 50){$x$}
\end{picture}
\caption{The graph of $\frac{U(x)}{U_0}$ for $K_1=3, \g=3, \l =1, \e =10$.
There exists a non-supersymmetric AdS minimum. \label{f1}}
\vspace{3cm}
\epsfxsize=4in \epsffile{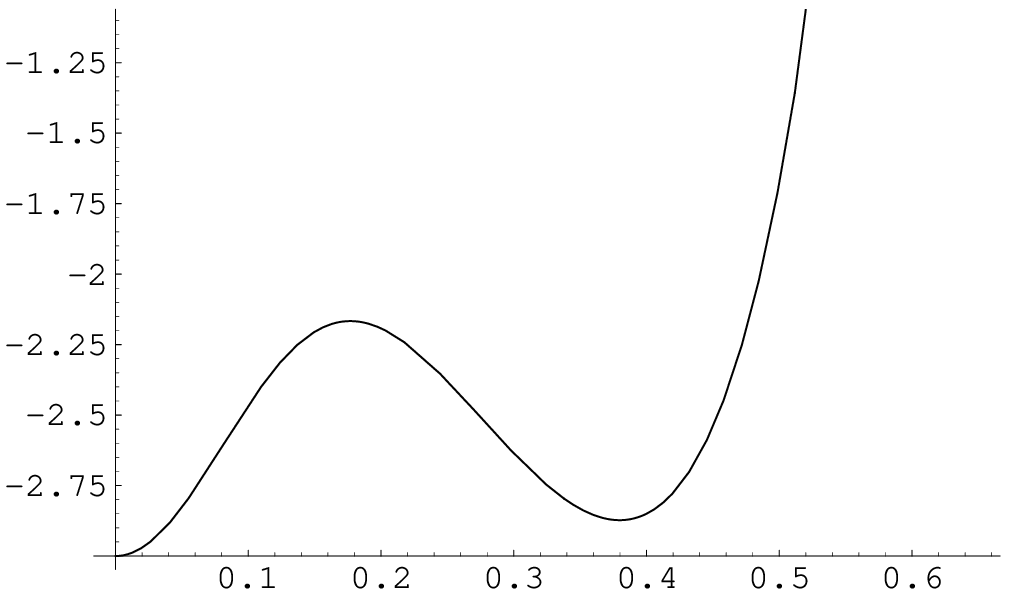}
\begin{picture}(30,30)
\put(-255,185){$\frac{U(x)}{U_0}$}
\put(5, 15){$x$}
\end{picture}
\caption{The graph of $\frac{U(x)}{U_0}$ for $K_1=5, \g=2.5, \l =2, \e =10$.
There exists a non-supersymmetric AdS minimum. \label{f3}}
\end{figure}

As an approximation, we will assume that the presence of this extra five-brane will not 
modify much the vacuum constructed in the previous section.
As a result, we can talk about the effective potential $U(x)$
describing dynamics of the five-brane. 
In fact, it is possible to show that the vacuum value of the moduli $S, T^I, {\bf Y}$
receive corrections of order $x^2$. Therefore, for very small values of $x$, $x << 1$, their vacuum 
values will not shift much. This suggest that the effective potential $U(x)$ is 
a decent approximation. Of course, in order to describe the system exactly,
one has to solve all the equations for moduli including the equations
for the imaginary parts. This is not possible to do analytically.
However, it is natural to argue that the qualitative behavior of this system
will be captured assuming that there is the effective potential $U(x)$ with
the rest of the moduli fixed along the lines of the discussion in the previous
section.
Thus, we consider dynamics of one field ${\bf X}$ with the Kahler potential
\begin{equation}
\frac{K({\bf X})}{M^2_{Pl}}=K_0+\frac{1}{4}K_1 ({\bf X}+\bar {\bf X})^2 
\label{2.6}
\end{equation}
and the superpotential
\begin{equation}
W({\bf X})=W_0-W_1 e^{-\e \g  {\bf X}^2},
\label{2.7}
\end{equation}
where $K_0$ is a constant independent of ${\bf X}$, $K_1$ is given by
\begin{equation}
K_1=\frac{2 \tau^{'}_5}{VRb^2},
\label{2.8}
\end{equation}
$W_0$ is a constant of order fluxes, 
\begin{equation}
W_0 \sim W_f,
\label{2.9}
\end{equation}
$W_1$ is approximately given by (see eq.~\eqref{1.46})
\begin{equation}
W_1 =\frac{F_1}{\e} W_f =\frac{F_1}{\e}W_0
\label{2.10}
\end{equation}
and the coefficient $\g$ is given by 
\begin{equation}
\g=\frac{\b^{'}}{T^2}.
\label{2.11}
\end{equation}
Without loss of generality, we can set $ImT^2=0$. Then 
\begin{equation}
\g=\frac{\b^{'}}{Rb^2}.
\label{2.12}
\end{equation}
The effective potential for the ${\bf X}$ modulus is given by
\begin{equation}
U({\bf X})=e^{\frac{K({\bf X})}{M^2_{Pl}}}
(G^{-1}_{{\bf X}\bar {\bf X}} D_{{\bf X}}W({\bf X})D_{\bar {\bf X}}\bar W(\bar {\bf X})
-3W({\bf X})\bar W(\bar {\bf X})),
\label{2.13}
\end{equation}
where the Kahler covariant derivative is defined as usual
\begin{equation}
D_{{\bf X}}W({\bf X})= \partial_{{\bf X}} W({\bf X})+\frac{1}{M^2_{Pl}}
\partial_{{\bf X}} K({\bf X})W({\bf X}).
\label{2.14}
\end{equation}
As was argued before, the imaginary part of ${\bf X}$ can be stabilized 
by this potential. Therefore, the potential $U({\bf X})$ can be treated
as an effective potential for one real field $x$.
We will denote it $U(x)$. From eqs.~\eqref{2.6} and~\eqref{2.7}, 
we obtain
\begin{equation}
U(x)=U_0 e^{K_1x^2}(-3+2K_1x^2(1+ \l e^{-\e \g x^2})^2), 
\label{2.15}
\end{equation}
where $U_0$ is a constant of order $\frac{W_f^2}{M^2_{Pl}}$,
$K_1$ and $\g$ are given by eqs.~\eqref{2.8} and~\eqref{2.12} respectively
and $\l$ is given by
\begin{equation}
\l =\frac{2 \g F_1}{K_1}.
\label{2.16}
\end{equation}
Eq.~\eqref{2.15} gives an effective potential $U(x)$. 
Unfortunately, it is very difficult to analyze 
this potential analytically. A graphical 
analysis shows that, generically, 
this potential has a non-supersymmetric AdS vacuum for a non-zero value of $x$.
The form of $U(x)$ for various choices of parameters is shown on 
Figures~\ref{f1} and~\ref{f3}. 
It is possible to adjust parameters in such a way that 
the vacuum becomes dS. However, in this case, the parameter $\l$
has to be taken to be sufficiently greater than one, whereas eqs.~\eqref{2.16}, \eqref{2.8} and~\eqref{2.12}
require that $\l$ be of order one. Therefore, for reasonable values of the parameters,
the minimum is always AdS. It is possible to adjust parameters so that $x$ 
is less than the size of the interval, which, as discussed
in the previous section, can be stabilized at a value of order one.
%
%
%
%
%
%
%
This AdS vacuum can be raised to a metastable 
dS vacuum by methods discussed in~\cite{Raise}. 
This demonstrates that the most general system of heterotic M-theory 
moduli can be stabilized in a dS vacuum.

In the rest of the paper, we will be interested
in dynamics of a five-brane in the regime $x<<1$. 
Heterotic M-theory vacua can contain
several five-branes wrapped on non-isolated genus zero or higher
genus curves. We have just argued that those five-branes 
which are located sufficiently far away from the visible brane 
can be stabilized. Now we would like to understand the fate 
of the five-branes which are close to the visible sector.
Such five-brane will roll towards $x=0$. 
The potential $U(x)$ in this regime does not lead to any interesting physics.
It does not provide stabilization of $x$. It is also hard to imagine how to use it
in any cosmological framework. On the one hand, it is negative and, hence, cannot 
be used for inflation. On the other hand, it does not satisfy conditions necessary
for Ekpyrotic cosmology~\cite{Justin1, Justin2, Justin3}.
To make use of this potential, we will modify it by Fayet-Iliopoulos terms. 
Depending on relations among various coefficients,
Fayet-Iliopoulos terms can lead to either stabilization 
of $x$ or a potential with certain inflationary properties.


\subsection{Fayet-Iliopoulos Terms and dS Vacua}


In both weakly and strongly coupled heterotic string models, there can be anomalous 
$U(1)$ gauge groups. They can arise in both the visible and the hidden sectors.
The anomaly is canceled by a four-dimensional version of the Green-Schwarz 
mechanism. This anomalous $U(1)$ gives rise to the Fayet-Iliopoulos term~\cite{DSW}, 
which, in turn, gives rise to the 
moduli effective potential
of the form
\begin{equation}
U_D =M_{Pl}^4 g^2 \frac{b}{V^2}, 
\label{2.17}
\end{equation}
where $b$ is a constant and $g$ is the gauge coupling constant. In the context of the strongly 
coupled heterotic string theory, the coefficient $b$ was estimated in~\cite{Raise}
and was found to be, generically, of order
\begin{equation}
b \sim 10^{-18}.
\label{2.18}
\end{equation}
The potential $U_D$ depends on in what sector there appears an anomalous $U(1)$. 
The reason is that the coupling constants in the visible and the hidden sectors are different.
They are given by~\cite{Nonstandard}
\begin{equation}
g_{visible}^2 =\frac{g^2_0}{Re(S+\alpha_1^{(1)}T^1+ \alpha_2^{(1)}T^2+
\b (T^1-\frac{{\bf Y}^2}{T^1})+ \b^{'} (T^2-\frac{{\bf X}^2}{T^2}) )}
\label{2.19}
\end{equation}
and 
\begin{equation}
g_{hidden}^2 =\frac{g^2_0}{Re(S-\alpha_1^{(2)}T^1-\alpha_2^{(2)}T^2 +
\b\frac{{\bf Y}^2}{T^1}+\b^{'}\frac{{\bf X}^2}{T^2})},
\label{2.20}
\end{equation}
where $g_0$ is a moduli independent constant of order $\alpha_{GUT}$. 
In~\cite{BKQ, Raise}, Fayet-Iliopoulos potentials $U_D$ were 
used to raise AdS vacua to dS vacua. In this paper, we will be
interested in the $x$ dependence of $U_D$. If the anomalous $U(1)$ 
appears in the hidden sector, the potential $U_D(x)$ takes the form
\begin{equation}
U^{visible}_D(x)=\frac{B_1}{B_2-\g x^2}, 
\label{2.21}
\end{equation}
whereas, 
if the anomalous $U(1)$ appears in the hidden sector, the potential $U_D$ 
is
\begin{equation}
U^{hidden}_D(x)=\frac{C_1}{C_2+\g x^2},
\label{2.22}
\end{equation}
where, $B1, B2, C1$ and $C_2$ can be read off from
eqs.~\eqref{2.17}, \eqref{2.19} and~\eqref{2.20} and $\g$ is given by 
eq.~\eqref{2.12}. 

\begin{figure}
\epsfxsize=4in \epsffile{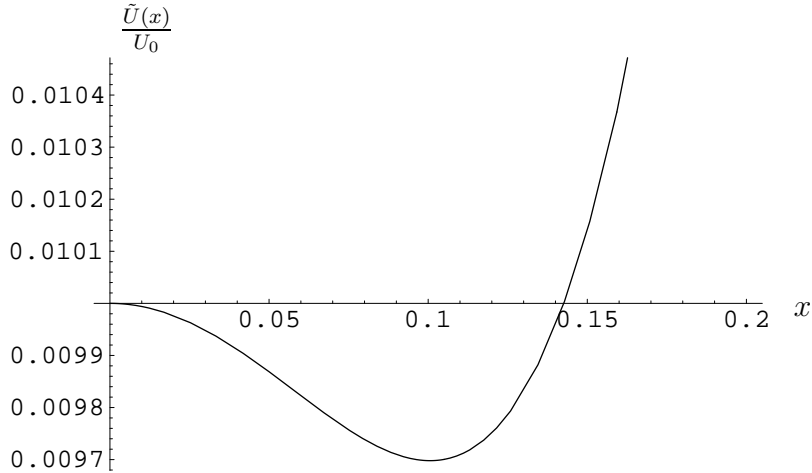}
\begin{picture}(30,30)
\put(-250,165){$\frac{\tilde{U}(x)}{U_0}$}
\put(5, 60){$x$}
\end{picture}
\caption{The graph of $\frac{\tilde{U}(x)}{U_0}$ in the regime $x<<1$
for $a_2=2.95, \g=1, \frac{C_1}{U_0} =3.01, C_2 =1$.
There exists a dS minimum. \label{f2}}
\end{figure}

We would like to modify our potential $U(x)$ by $U_{D}(x)$. 
In this section, we take $U_D(x)$
to be $U_D^{hidden}(x)$. We will now show that the potential energy 
\begin{equation}
\tilde{U}(x)=U(x)+U_{D}^{hidden}(x)
\label{2.23}
\end{equation}
can provide stabilization of $x$ in the regime $x<<1$ in a dS vacuum.
We should point out that, if we modify $U(x)$ by some
other moduli dependent correction, it is not 
very obvious that this correction will not destabilize
other, additional to $x$, moduli. However,
in~\cite{Raise}, it was shown that if the order 
of magnitude of $U_D$ is the same as (or less than) the order 
of magnitude of $U$, it is possible 
to find a solution to eqs. $d(U+U_D)=0$ fixing all the moduli
considered in the previous section. Therefore, it is still a
decent approximation to consider the effective potential
$\tilde{U}(x)$ assuming that all the remaining moduli are fixed.
Now note the following simple facts. 
Since $x=0$ is the minimum of the function $U$, for small $x$ we have
\begin{equation}
\frac{\partial U(x)}{\partial x} >0
\label{2.24}
\end{equation}
On the other hand, from eq.~\eqref{2.22}, it follows that
\begin{equation}
\frac{\partial U_D^{hidden}(x)}{\partial x} <0.
\label{2.25}
\end{equation}
This means that it should be possible to find a solution to the equation
\begin{equation}
\frac{\partial \tilde{U}(x)}{\partial x} =0
\label{2.26}
\end{equation}
under mild assumptions. 
For $x<<1$, the potential $U(x)$ is governed by the quadratic function
\begin{equation}
U(x)=-3U_0+a_2 U_0 x^2,
\label{2.26.1}
\end{equation}
where $a_2$ is given by
\begin{equation}
a_2=K_1(2(1+\l)^2-3).
\label{2.28}
\end{equation}
%
Using eqs.~\eqref{2.8}, \eqref{2.12}, \eqref{2.16} and~\eqref{1.47},
one can
show that $a_2$ is greater than zero for any choice of the parameters.
It is straightforward to solve eq.~\eqref{2.26} in this regime.
The approximate solution is
\begin{equation}
x_{min} \approx
\frac{1}{\sqrt{\g}}\left(\sqrt{\frac{\g C_1}{a_2U_0}} -C_2\right)^{1/2},
\label{2.27}
\end{equation}
It is possible to adjust the parameters so that $x_{min}$ is real and much less than one.
It is also straightforward to show that, if the solution for $x_{min}$ exists, it is always a minimum.
The simplest way to do it is prove that, if the solution~\eqref{2.27} exists, then
$x=0$ is always a maximum. Since $x_{min}<<1$, the value of the cosmological constant
is approximately given by
\begin{equation}
\Lambda \approx -3U_0 +\frac{C_1}{C_2}.
\label{2.29}
\end{equation}
It is obvious that $\Lambda$ can be of both signs. By fine-tuning it is possible
to set 
\begin{equation}
\Lambda \sim 10^{-120}M_{Pl}^4
\label{2.30}
\end{equation}
which is consistent with observations. The form of the potential $\tilde{U}(x)$ in the
regime $x<<1$ 
is shown on Figure~\ref{f2}.

In Sections 2 and 3, we showed that the most general system of heterotic M-theory moduli can be 
stabilized in a dS vacuum. In addition to moduli considered in~\cite{Raise}, 
we also provided stabilization for extra $h^{1,1}$ moduli and an extra five-brane
associated with a non-isolated genus zero curve or with a higher genus curve.
In the presence of such a five-brane,
the system of moduli
can be stabilized by fluxes and non-perturbative effects
in a non-supersymmetric AdS vacuum which then can be lifted to a dS vacuum
as in~\cite{Raise}. This five-brane can also 
be stabilized by balancing
the supergravity potential energy against a Fayet-Iliopoulos term induced by an anomalous 
$U(1)$ gauge group in the hidden sector.
Thus, the potential energy $\tilde{U}(x)$ might admit two dS vacua.
One of the them is the lift of the non-supersymmetric AdS vacuum.
The other one can additionally arise for $x <<1$, though 
it did not existed in the absence of the Fayet-Iliopoulos term.


\section{The Five-Brane Modulus as an Inflaton}


\subsection{Constructing an Inflationary Potential}


We begin this section with modifying the potential energy $U(x)$ by 
the Fayet-Iliopoulos term $U_{D}^{visible}(x)$ given by~\eqref{2.21}.
The first derivative of $U_{D}^{visible}(x)$ is positive, hence, 
the potential
\begin{equation}
\tilde{U}(x)=U(x)+U_{D}^{visible}(x)
\label{3.1}
\end{equation}
does not have a non-trivial minimum for $x <<1$ and $x$ rolls towards $x=0$. 
We will assume that the potential $\tilde{U}(x)$ is positive.
The potential~\eqref{3.1} has the following form
\begin{equation}
\tilde{U}(x)=U_0 e^{K_1x^2}(-3+2K_1x^2(1+\l e^{-\e \g x^2})^2) +\frac{B_1}{B_2-\g x^2}.
\label{3.2}
\end{equation}
Let us recall that the coefficients $\e, K_1, \g$ and $\l$ are given by 
eqs.~\eqref{1.21}, \eqref{2.8}, \eqref{2.12} and~\eqref{2.16}, 
$U_0$ is a constant of order $\frac{W_f^2}{M_{Pl}^2}$ and 
$B_1$ and $B_2$ can be read off from eqs.~\eqref{2.17} and~\eqref{2.19}.
We assume that this potential is positive, that is, 
\begin{equation}
\frac{B_1}{B_2} >3U_0.
\label{3.3}
\end{equation}
%
Our goal is to examine whether this potential can satisfy
the slow roll conditions required by inflation.
As in the previous section, we are interested in the regime
$x <<1$. Then we can expand $\tilde{U}(x)$ in powers of $x$. For 
our purposes, it is enough to keep only two leading terms.
We obtain
\begin{equation}
\tilde{U}(x)\approx A_0+A_2x^2,
\label{3.4}
\end{equation}
where 
\begin{equation}
A_0 =-3U_0 +\frac{B_1}{B_2}
\label{3.5}
\end{equation}
and
\begin{equation}
A_2=K_1(2(1+\l)^2-3)U_0 +\frac{\g B_1}{B_2^2}=a_2+\frac{\g B_1}{B_2^2}.
\label{3.6}
\end{equation}
%
In order to study the standard slow roll parameters $\e (x)$ and $\eta (x)$, 
we have to canonically normalize the kinetic energy. 
From the Kahler potential~\eqref{2.6}, it follows
that we have to redefine $x$ as 
\begin{equation}
x \to \sqrt{\frac{2}{K_1}}\frac{x}{M^2_{Pl}}.
\label{3.7}
\end{equation}
This new $x$ is canonically normalized and has dimension one. 
The potential energy now looks as follows
\begin{equation}
\tilde{U}(x) = A_0 +\frac{2 A_2}{K_1 M^2_{Pl}} x^2.
\label{3.8}
\end{equation}
To have inflation as $x$ rolls towards $x=0$, the two parameters 
\begin{equation}
\e (x) =\frac{M^2_{Pl}}{2} \left( \frac{\tilde{U}^{'}(x)}{\tilde{U}(x)}\right)^2
\label{3.9}
\end{equation}
and 
\begin{equation}
\eta (x) =M^2_{Pl} \frac{\tilde{U}^{''}(x)}{\tilde{U}(x)}
\label{3.10}
\end{equation}
have to be much less than one. 
From eq.~\eqref{3.8} we obtain 
\begin{equation}
\e (x) =\frac{2 A_2^2}{A_0^2 K_1^2 M_{Pl}^2}x^2.
\label{3.11}
\end{equation}
Clearly, for $x <<M_{Pl}$, $\e(x)$ is naturally much less than one. 
For the parameter $\eta(x)$ we have
\begin{equation}
\eta (x) =\frac{4A_2}{K_1 A_0}.
\label{3.12}
\end{equation}
Therefore, we need to impose
\begin{equation}
\frac{4}{K_1}A_2 < A_0.
\label{3.13}
\end{equation}
Using eqs.~\eqref{3.5} and~\eqref{3.6}, this condition can be rewritten as 
\begin{equation}
\frac{4(2(1+\l)^2-3)U_0 +\frac{4 \g B_1}{K_1 B_2^2}}{-3U_0 +\frac{B_1}{B_2}} <<1.
\label{3.14}
\end{equation}
The only way this can be fulfilled is when
\begin{equation}
\frac{B_1}{B_2} >> U_0
\label{3.15}
\end{equation}
and 
\begin{equation}
\frac{4 \g}{K_1 B_2} << 1.
\label{3.16}
\end{equation}
Condition~\eqref{3.15} is a relatively mild constraint.
Using eqs.~\eqref{2.8}, \eqref{2.11}, \eqref{2.17} and~\eqref{2.19}, condition~\eqref{3.16}
can be rewritten as
\begin{equation}
\frac{2V}{Re (S+\alpha_1^{(1)}T^1+ \alpha_2^{(1)}T^2+
\b (T^1-\frac{{\bf Y}^2}{T^1})+ \b^{'} T^2)} <<1.
\label{3.17}
\end{equation}
Unfortunately, it does not seem to be possible to fulfill 
this condition, at least in the context of low-energy field theory.
Inequality~\eqref{3.17} requires that some of the tensions 
$\alpha_I^{(1)}, \b$ or $\b^{'}$ be much greater than one. In this case, 
one cannot trust, even approximately, expressions~\eqref{2.19} and~\eqref{2.20}
for the coupling constants because they can be substantially modified 
by higher order corrections~\cite{Curio1, Curio2}. 
On the other hand, eq.~\eqref{3.17} may make perfect
sense in the context of M theory. 
However, we would like to stay within the context of low-energy 
field theory. All we have to do to make the parameter $\eta(x)$ small is to decrease 
the parameter $A_2$ in eq.~\eqref{3.4}. This, in fact, can easily be done. We just have to 
replace a Fayet-Iliopoulos term in the visible sector by a Fayet-Iliopoulos term in the 
hidden sector. Equivalently, we could just add the Fayet-Iliopoulos term in the 
hidden sector to eq.~\eqref{3.1}. In both cases, addition of such a Fayet-Iliopoulos term
increases $A_0$ and decreases $A_2$. 
It is possible to (not necessarily fine) tune 
the parameter $\eta$ to be much less than one. 
Let us consider it in slightly more detail.
Assuming, for simplicity, that only a hidden sector Fayet-Iliopoulos term 
is present and using~\eqref{2.22}, 
we have
\begin{equation}
\eta (x) =
\frac{4(2(1+\l)^2-3)U_0 -\frac{4 \g C_1}{K_1C_2^2}}
{-3U_0+\frac{C_1}{C_2}} <<1.
\label{3.17.1}
\end{equation}
We can rewrite~\eqref{3.17.1} as
\begin{equation}
\eta (x) =
\frac{4(2(1+\l)^2-3-\frac{3\g}{K_1C_2})U_0 -\frac{4 \g }{K_1C_2}A_0}{A_0} <<1.
\label{3.17.2}
\end{equation}
Since the quantities $U_0$ and $A_0$ are, generically, of the same order of 
magnitude~\cite{Raise}, ineq.~\eqref{3.17.2} is a relatively mild constraint.
In the next subsection, we will show that the parameter $\eta$ does
not have to be fine tuned to be very small.
As discussed in the previous section,
addition of a Fayet-Iliopoulos term in the 
hidden sector can stabilize $x$. 
This happens if the numerator in~\eqref{3.17.2} becomes negative.
In this case, the point $x=0$ becomes a maximum and 
the potential $\tilde{U}(x)$ acquires a minimum 
at a non-zero value of $x$. 
This was studied in the previous section.
In this section, we assume that
the effect of such an addition is to make the potential flat,
rather than to produce a non-trivial minimum.

In this subsection, we showed that the five-brane effective potential, 
with various Fayet-Iliopoulos terms included,
can satisfy the slow roll conditions
\begin{equation}
\e (x) <<1, \quad \eta (x) << 1
\label{3.18}
\end{equation}
necessary for inflation.
Let us recall that the system of fields contains various other moduli,
in addition to $x$. The potential energy, by construction, 
has a minimum in these directions. Therefore, dynamically, one 
expects that they will roll fast in the minimum, leaving 
the modulus $x$ to roll slowly. Since ineqs.~\eqref{3.18}
are satisfied for $x <<1$, the five-brane modulus $x$ can be 
viewed as an inflaton. 


\subsection{The Amount of Inflation and Primordial Fluctuations}


In this subsection, we will consider the amount of inflation and 
primordial fluctuations. The amount of inflation is defined by
\begin{equation}
N =\ln \frac{a_f}{a_i},
\label{3.20}
\end{equation}
where $a_i$ and $a_f$ are the initial and final values of the expansion 
parameter. The evolution of $a$ and $x$ can be found from
the Friedmann equation
\begin{equation}
H^2 =\frac{1}{3 M^2_{Pl}}(\frac{1}{2} \dot x^2 +\tilde{U}(x))
\label{3.21}
\end{equation}
and the $x$-equation of motion
\begin{equation}
\stackrel{\cdot \cdot}{x} +3H \dot x +\tilde{U}^{'}(x)=0,
\label{3.22}
\end{equation}
where 
\begin{equation}
H=\frac{\dot a}{a}
\label{3.23}
\end{equation}
is the Hubble constant.
Since during the period of inflation the kinetic energy is much less than the potential energy, 
it follows from eq.~\eqref{3.24} that
\begin{equation}
H \approx \frac{1}{M_{Pl}} \sqrt{\frac{A_0}{3}}.
\label{3.24}
\end{equation}
This gives 
\begin{equation}
a(t) \approx a_i e^{\sqrt{\frac{A_0}{3 M^2_{Pl}}} t}.
\label{3.25}
\end{equation}
Similarly, integrating eq.~\eqref{3.22} we find that
\begin{equation}
x(t) \approx x_i e^{-\frac{4 A_2}{K_1 M_{Pl}\sqrt{3 A_0}}t}.
\label{3.27}
\end{equation}
From eqs.~\eqref{3.25} and~\eqref{3.27} we obtain
\begin{equation}
N = \ln \frac{a_f}{a_i} = \frac{A_0 K_1}{4A_2}
\ln \frac{x_i}{x_f} =\frac{1}{\eta} \ln \frac{x_i}{x_f}, 
\label{3.28} 
\end{equation}
where eq.~\eqref{3.12} has been used.
By $x_i$ and $x_f$ we denoted the initial and final positions 
of the five-brane during inflation. Taking, as an an example,
\begin{equation}
\eta \sim 0.1, \quad \frac{x_i}{x_f} \sim 10^4.
\label{3.29}
\end{equation}
we get 
\begin{equation}
N \sim 80
\label{3.30}
\end{equation}
which is consistent with observations. 
Primordial fluctuations are determined by the following quantity
\begin{equation}
\delta^2_H =\frac{4}{25} 
\left(\frac{H}{\dot x} \right)^2
\left(\frac{H}{2 \pi} \right) =
\frac{1}{150 \pi^2} \frac{\tilde{U}(x)}{M^4_{Pl}\e(x)}
\approx \frac{1}{150 \pi^2} \frac{A_0}{M^4_{Pl}\e(x)},
\label{3.31}
\end{equation}
where, in our case, $\e (x)$ is given by eq.~\eqref{3.11}.
Note that as $x$ goes to zero, $\e (x)$ goes to zero and $\delta^2_H$ 
goes to infinity. Therefore, it is important to terminate 
inflation before the fluctuations became too big. This will be 
discussed in the next subsection. Taking the order of magnitude of 
$A_0$ set by the fluxes (see eq.~\eqref{1.19} and discussion below it),
\begin{equation}
A_0 \sim M_{Pl}^4 10^{-18},
\label{3.32}
\end{equation}
$\e (x)$  to be
\begin{equation}
\e (x) \sim 10^{-12},
\label{3.33}
\end{equation}
corresponding, for example, to
\begin{equation}
\eta \sim 0.1, \quad \frac{x_f}{M_{Pl}} \sim 10^{-5},
\label{3.34}
\end{equation}
we obtain
\begin{equation}
\delta^2_{H} \sim 10^{-10},
\label{3.35}
\end{equation}
which is consistent with measurements of the cmb anisotropy.

Thus, this model of inflation gives appropriate values for the
amount of inflation and primordial fluctuations. However, these
results really make sense only if it is possible to escape from inflation 
before the fluctuations became too big.


\subsection{Escape from Inflation}


At very small values of $x$, we cannot really trust the potential $\tilde{U}(x)$
because one should expect extra light states to become light as we approach the 
singularity $x=0$. At the present time, the new physics at distances much less than the 
eleven-dimensional Planck scale is not known. It may happen that these 
new states are string-like, rather than particles~\cite{Hanany}.
In this subsection, we would like to give a qualitative argument 
how such new states can terminate inflation. Let us emphasize that 
we cannot prove that this is the actual mechanism. We just 
would like to point out that the appearance of new physics 
at short distances can help to terminate inflation.
We will assume that the new states are particles and, in the 
absence of fluxes and non-perturbative effects, the moduli space 
of heterotic M-theory is describable by the superpotential
\begin{equation}
{\mathbb W}={\mathbb W}(\Phi, {\bf X}), 
\label{3.36}
\end{equation}
where the fields $\Phi$ come from a membrane stretching between 
the visible brane and the five-brane. These fields are expected to be 
charged under $E_8$. 
The moduli space, that is the space of solutions of eq.
\begin{equation}
d {\mathbb W}=0,
\label{3.37}
\end{equation}
must consist of two branches. The first branch, the five-brane branch, is characterized by a non-zero expectation value
of the five-brane translational modulus $x$. In this branch, the five-brane multiplet is massless, while 
the fields $\Phi$ are massive and integrated out form the low-energy field theory.
The mass of the fields $\Phi$ is proportional to $x$. 
The second branch, the instanton branch, is characterized by the vanishing expectation value of $x$
and coincides with the moduli space of transition moduli~\cite{BDO1} of an instanton on our  
Calabi-Yau threefold. This five-brane-instanton transition is called small instanton 
transition~\cite{Seiberg, KS, OPP, BDO1}. The interpretation of the transition is the following.
As the five-brane hits the visible brane, it changes the vector bundle on the Calabi-Yau 
manifold. The second Chern class of the new vector bundle changes by the amount 
of the curve on which the five-brane was wrapped. This new bundle has more moduli.
The new moduli are precisely the transition moduli parameterizing the instanton 
branch of the superpotential ${\mathbb W}$. In the instanton branch, only those 
components of $\Phi$, which correspond to the transition moduli, take non-zero expectation 
values and are massless, the remaining ones become massive and get integrated out.
The five-brane is also massive and integrated out in the instanton branch.
The origin represents a singularity, where all the multiplets become massless.
From the bundle viewpoint, the singularity at the origin corresponds to a vector bundle 
becoming singular and turning into a torsion free sheaf~\cite{OPP}. 
From the five-brane view-point, the singularity at the origin
corresponds to a five-brane coinciding with the visible brane.
An analogous, but simpler, transition takes place in string theory in
the $Dp-D(p+4)$ system~\cite{Douglas}.
The $Dp-D(p+4)$ system is describable by a supersymmetric field theory with eight
supercharges. The moduli space of this system consists of two branches, the Coulomb 
branch and the Higgs branch. The Coulomb branch describes positions of the
$Dp$-brane away from the $D(p+4)$-brane. The Higgs branch describes how 
the $Dp$-brane can get dissolved into the $D(p+4)$-brane. Geometrically, the Higgs branch 
is isomorphic to the one-instanton moduli space on a $4$-manifold which is just the ADHM moduli space.
In the heterotic M-theory case, the analog of the Coulomb branch is the moduli space 
of five-branes. The analog of the Higgs branch is the space of transition moduli.
Thus, transition moduli can be understood as a Calabi-Yau threefold generalization
of the ADHM one-instanton moduli space.

Now let us see how this picture changes in the presence of fluxes 
and non-perturbative effects. For $x$ very small, we have to include
$\Phi$ into the Lagrangian. The Kahler potential receives an extra contribution
\begin{equation}
K_{\Phi}\sim tr_{E_8} (\Phi \bar \Phi).
\label{3.39}
\end{equation}
Then it is not hard to show that the potential energy 
will contain, among others, a term 
\begin{equation}
-\frac{U_0}{M^2_{Pl}} tr_{E_8} (\Phi \bar \Phi).
\label{3.40}
\end{equation}
This is a consequence of the following very simple statement.
If we take the theory with the constant superpotential and quadratic
Kahler potential, the potential energy has a maximum 
at zero and for small fields is dominated by a negative 
quadratic term.
Therefore, the mass of the fields $\Phi$ schematically behaves as
\begin{equation}
|\frac{\partial^2{\mathbb W}(\Phi, {\bf X})}{\partial \Phi^2}\mid_{\Phi =0}|^2
- \frac{U_0}{M^2_{Pl}}.
\label{3.41}
\end{equation}
This means that as the five-brane gets very close to the visible sector, 
the fields $\Phi$ become tachyonic and begin to roll downhill.
Since no slow roll conditions on $\Phi$ are satisfied, this terminates
inflation. 
Eventually, the five-brane hits the wall and disappears (gets massive and integrated out
from the low-energy field theory) through a small instanton transition. 
In addition, those components 
of $\Phi$, which do not correspond to the transition instanton moduli, get 
a mass, according to the superpotential~\eqref{3.36}, and get integrated out.
Now their mass is set by the vacuum expectation values of the transition moduli.
The new system of moduli after the small instanton transition
involves the moduli discussed in Section 2 plus the new transition moduli.
The physics of them will be considered in the next section.

This escape from inflation is, somewhat, analogous to one in 
$D3$-$D7$ inflation studied in~\cite{Dasgupta}.
In~\cite{Dasgupta}, the negative mass for the charged hypermultiplets
(analogs of our fields $\Phi$) was created by Fayet-Iliopoulos terms. 
The same Fayet-Iliopoulos terms were responsible for stabilization
at a non-zero value. 


\section{The Post-inflationary Phase}


After inflation, the five-brane disappears through a small 
instanton transition and turns into new vector bundle moduli, which we 
denote by $\phi_i$. The new system of moduli contains now the following fields
\begin{equation}
S, T^1, T^2, {\bf Y}, Z_{\alpha}, \phi_i.
\label{3.42}
\end{equation}
This system of moduli can be stabilized. The moduli 
$S, T^1, T^2, {\bf Y}$ and $Z_{\alpha}$ can be stabilized by the same 
mechanism as in Section 2. In this section, we will 
concentrate on the moduli $\phi_i$. Without loss of generality, we 
can assume that there is only one such modulus $\phi$.
Vector bundle moduli have a non-perturbative superpotential.
It appears as a factor in eqs.~\eqref{1.35}, \eqref{1.36} and~\eqref{1.37}.
Since after the small instanton transition only the bundle on the visible brane has
changed, only $W_{v5}$ and $W_{vh}$ will depend on $\phi$ after the transition.
In fact,
\begin{equation}
W_{v5} >>W_{vh},
\label{3.43}
\end{equation}
since the coefficient $\tau$ in eqs.~\eqref{1.35} and~\eqref{1.36} 
is much greater than one. Therefore, the potential for $\phi$ is 
\begin{equation}
W(\phi)={\cal P}(\phi)e^{-\tau {\bf Y}},
\label{3.44}
\end{equation}
where we have denoted the factor depending on $\phi$ by ${\cal P}(\phi)$.

Let us make a remark. One can worry that, after the transition, the superpotential~\eqref{3.44} 
will not depend on $\phi$. Indeed, the superpotential is induced by a string wrapping
isolated genus zero curves. On the other hand, the five-brane that turned into the modulus $\phi$ 
was wrapped on a non-isolated genus zero curve or a higher genus curve. 
It seems possible that the bundle over the isolated curves did not really change
and the bundle moduli contribution to the superpotential will remain unchanged.
However, this logic is not quite correct because the curves might intersect.
If the curve on which the five-brane was wrapped intersects at least one isolated
genus zero curve, the non-perturbative superpotential will 
change and it will depend on $\phi$.

Even though the five-brane modulus $x$ could not be stabilized, the new vector
bundle modulus $\phi$ can.
The equation 
\begin{equation}
D_{\phi}W=0
\label{3.45}
\end{equation}
has a solution. The analysis of this equation 
will be analogous to the one in~\cite{BO}.
Let us recall that the superpotential $W$ is a sum
of various contributions
\begin{equation}
W=W_f+W_{g}+W_{np},
\label{3.46}
\end{equation}
where $W_{np}$ is now the sum of $W_{5h}$ and $W(\phi)$. 
Then eq.~\eqref{3.45} can be written as 
\begin{equation}
\partial_{\phi}{\cal P}(\phi) e^{-\tau {\bf Y}}=
\partial_{\phi}K(\phi) W_f,
\label{3.47}
\end{equation}
where we have assumed that
\begin{equation}
W_f >> W_{\phi}.
\label{3.48}
\end{equation}
In~\cite{BO}, it was shown that ineq.~\eqref{3.48} is indeed satisfied.
Vector bundle moduli superpotentials were studied in detail in~\cite{BDO2, BDO3}.
They were found to be high degree polynomials.
Therefore, we will take 
\begin{equation}
{\cal P}(\phi)=\phi^d.
\label{3.49}
\end{equation}
The Kahler potential $K(\phi)$ represents a more complicated problem.
It was evaluated explicitly only for bundles which can be written
as the pullback of a bundle on a four-manifold~\cite{Gray}.
A generic bundle on a threefold is, obviously, not of this form.
In this paper, we will not need the actual form of the Kahler 
potential. Using eq.~\eqref{3.49} and ignoring the imaginary part of ${\bf Y}$,
eq.~\eqref{3.47}
can be written as 
\begin{equation}
\phi^d e^{-\tau y}=\partial_{\phi}K(\phi) W_f.
\label{3.50}
\end{equation}
Clearly, such an equation has a solution for a generic function $K(\phi)$.
To present a numeric solution, we need to know the order of magnitude 
of $K$. It was estimated in~\cite{BO} and found to be $\sim 10^{-5}$
in Planck units.
Then, if $d$ is sufficiently large, we can approximately write eq.~\eqref{3.50} as follows
\begin{equation}
\phi^d \sim 10^{-5} e^{\tau y} W_f.
\label{3.51}
\end{equation}
Taking, as an example, 
\begin{equation}
W_f \sim 10^{-10}, \quad \tau \sim 200, \quad y \sim 0.8, \quad d \sim 40, 
\label{3.52}
\end{equation}
we obtain
\begin{equation}
\phi \sim 20.
\label{3.53}
\end{equation}
The relatively large value of $\phi$ means that the gauge connection 
is spread out over the Calabi-Yau manifold, rather than 
sharply peaked over some curve. As long as we stay away 
from singularities in the moduli space of vector bundles,
any value of $\phi$ is acceptable. To be slightly more precise,
eq.~\eqref{3.53} is a solution for the absolute value of $\phi$.
The phase of $\phi$ can also be found form eq.~\eqref{3.47}~\cite{BO}.

Thus, we have shown that the new system of moduli has much simple 
stability properties. That is, it is possible to fix 
all moduli after inflation. Now we are going to argue that
the cosmological constant in the new vacuum can be positive and 
fine tuned to coincide with observations. Recall that the cosmological constant 
during inflation is given by three contributions
\begin{equation}
\Lambda_{inflation}=-\Lambda_{SUGRA} +\Lambda_{D}^{hidden}+
\Lambda_{D}^{visible}.
\label{3.54}
\end{equation}
The first term in eq.~\eqref{3.54} is the contribution 
from the supergravity potential energy. It is negative and approximately 
given by $-\frac{W_f^2}{M_{Pl}^2}$.
The second term comes from the Fayet-Iliopoulos term
in the hidden sector. The last term is the contribution from 
the Fayet-Iliopoulos term 
in the visible sector. They second and the third terms are positive. 
In the previous section, we argued that the existence of Fayet-Iliopoulos terms 
in the visible sector is not relevant for inflation. 
Nevertheless, we will assume that they are present. 
The cosmological constant $\Lambda_{inflation}$ has to be positive and large.
After the small instanton transition, some of the contributions to the cosmological constant
might change, because they depend on the properties of the vector bundle.
The contribution $\Lambda_{SUGRA}$ might change because, as was argued in~\cite{GKLM}, 
the flux-induced superpotential may receive higher order corrections from Chern-Simons
invariants which depend on the choice of the bundle. 
The contribution $\Lambda_{D}^{visible}$ will change because 
it depends on the gauge connection. Moreover, since after a small 
instanton transition there is a possibility of changing the number 
of families of quarks and leptons~\cite{KS, OPP}, the corresponding 
$U(1)$ gauge group might not be anomalous anymore.  
In this case $\Lambda_{D}^{visible}$ will be zero.
However, $\Lambda_{D}^{hidden}$ will not change because the bundle 
on the hidden brane remains unchanged.
All these arguments show that, in principle, the cosmological constant changes after inflation. 
Since it receives both negative and positive contributions, 
it is possible, by fine tuning, to make it consistent with observations.


\section{Conclusion}

 
In this paper, we considered dynamics of the five-brane wrapped on 
a non-isolated genus zero or higher genus curve.
Non-perturbative superpotentials do not depend on moduli 
of such five-branes. 
We showed that fluxes and non-perturbative effects can stabilize such a five-brane 
in a non-supersymmetric AdS vacuum.
We also
showed
that addition of Fayet-Iliopoulos terms not only 
can raise this vacuum to a dS vacuum but also
can create one more dS vacuum.
We also stabilized $h^{1,1}$ moduli
which do not have non-perturbative superpotentials.
The cosmological constant of the vacuum can be positive and 
fine tuned to be consistent with observations.
This provides a generalization of results of~\cite{BO, Raise}
and shows that the most complete system of heterotic 
string moduli can be stabilized in a vacuum with a positive 
cosmological constant. In addition, we 
showed that, by modifying the supergravity 
potential energy with Fayet-Iliopoulos terms,
one can create an inflationary potential 
and treat the five-brane translational modulus 
as an inflaton. However, the potential cannot be trusted at very small 
distances because one should expect extra light states to appear.
We give a qualitative argument how such new states 
can terminate inflation. The idea is that, in the presence
of fluxes and non-perturbative effects, these extra states can become
tachyonic when the five-brane comes very close to the visible brane.
Eventually, the five-brane hits the visible brane.
The system undergoes a small instanton transition which changes the vector 
bundle on the visible brane.
The five-brane disappears from the low-energy spectrum
while the vector bundle moduli are created. They have simpler 
stability properties and, as a result, the new system 
of moduli can be stabilized. We also argue that the 
cosmological constant changes after the transition.
The cosmological constant after the transition can be 
fine tuned to be consistent with observations.


\section{Acknowledgements}


The author is indebted to Juan Maldacena
for lots of interesting and helpful discussions.
The work is supported by NSF grant PHY-0070928.


\end{document}